\documentclass[journal]{IEEEtran}
\usepackage{epsfig,graphics,subfigure,psfrag,amsmath,cases,placeins}
\usepackage{latexsym,amssymb,amsmath,epsfig,subfigure}

\author{Derrick Wing Kwan Ng,~\IEEEmembership{Student Member,~IEEE,}
        Vincent K. N. Lau,~\IEEEmembership{Senior Member,~IEEE}\vspace*{-5mm}
}

\title{Power Control and Performance Analysis of Outage-Limited Cellular Network with MUD-SIC and Macro-Diversity\thanks{Manuscript received October 20, 2008; revised January 5, 2010; accepted May 17, 2010. The review of this paper was coordinated by Prof. F. Santucci.
The authors are with the University of British Columbia and the Hong Kong University of Science and Technology, respectively. (e-mail: wingn@ece.ubc.ca; eeknlau@ee.ust.hk).}}
{\newtheorem{Thm}{Theorem}}
\newtheorem{Lem}{Lemma}

\newtheorem{Remark}{Remark}

\begin{document}
\maketitle

\begin{abstract}
In this paper, we analyze the uplink {\em goodput}
(bits/sec/Hz successfully decoded) and {\em per-user packet
outage} in a cellular network using multi-user detection with
successive interference cancellation (MUD-SIC). We consider non-ergodic fading channels where microscopic fading channel information is not available at the transmitters. As a
result, {\em packet outage} occurs whenever the data rate of
packet transmissions exceeds the instantaneous mutual
information even if powerful channel coding is applied for protection. We are
interested to study the role of {\em macro-diversity} (MDiv)
between multiple base stations on the MUD-SIC performance where
the effect of potential {\em error-propagation} during the SIC
processing is taken into account. While the jointly optimal power and decoding order in the MUD-SIC are NP hard problem, we derive a simple on/off power control and asymptotically optimal decoding order with respect to the transmit power.
Based on the information theoretical framework, we derive the
closed-form expressions on the total \emph{system goodput} as well as the
{\em per-user packet outage probability}. We show that the system
goodput does not scale with SNR due to mutual interference in the
SIC process and macro-diversity (MDiv) could alleviate the problem
and benefit to the system goodput.
\end{abstract}

\vspace*{-0.7cm}
\section{Introduction}
\label{sect:intro} \IEEEPARstart{T}here are two important technologies that could
substantially enhance the uplink performance of cellular systems,
namely the multi-user detection (MUD) and the macro-diversity
(MDiv). The MUD is effective to mitigate intra-cell interference
while the MDiv is effective to exploit of inter-cell interference
from adjacent base stations. It is well-known that jointly maximum likelihood
multi-user detection (ML MUD) is optimal but with exponential order
of complexity with respect to (w.r.t.) the number of users in the system. There are a lot
of research works on low complexity MUD such as the linear MUD
\cite{JR:Linear-MUD:Verdu,JR:Linear-MUD:ShengChen} and the
successive interference cancellation (SIC)
\cite{JR:SIC-analysis:94,CN:SIC-complexity}. In
\cite{Bettesh:00,JR:Tuninetti:02}, the authors analyzed the system
goodput (bit/s/Hz successfully delivered to mobile user) for
multi-access channels with minimum mean square error (MMSE)
detector. However, the MMSE MUD cannot achieve {\em Pareto
optimality} in the capacity region. On the other hand, MUD-SIC is a
promising technology at the base station to mitigate intra-cell interference at
reasonably low complexity. In this paper, we study the uplink
performance analysis of an outage-limited multi-cell system with
both MUD-SIC at each base station and MDiv between adjacent base stations. While there
are quite a number of works studying the MUD design and performance
analysis on single cell systems
\cite{book:multi-user-detection,JR:SIC-Khaled}, there are still a
number of open technical challenges to apply MUD-SIC in multi-cell
systems with MDiv. They are elaborated in the following:

\begin{itemize}
\item {\bf Per-user Outage and Error Propagation in MUD-SIC}
Conventional performance analysis of multi-access fading channel is
usually based on the ergodic capacity
\cite{JR:Shamai_1:97,JR:Shamai_2:97}. Uplink power adaptation for
multiaccess channel is addressed in \cite{Tse:98,CN:SIC-power-control,JR:SIC-imperfect-cancellation:03} where the transmit
power of mobile users are optimized with respect to a system
objective function of user capacities.
However, in all these works, they did not take into account of the
potential packet errors (and the error propagation effects in the
SIC process) due to channel outage. When error-propagation effect of
the MUD-SIC is considered, the packet error events between the $K$
users are coupled together and the outage event cannot be determined
by whether the rate vector is inside the capacity region or not\footnote{For
example, whether the 2nd decoded packet is successful depends not
only on the channel condition of that user but also on whether the
1st decoded packet is successful. Furthermore, even if a rate vector
is outside the multi-access capacity region, some user(s) may still
be able to decode the packet successfully. This substantially
complicated the analysis.}.

\item {\bf Power and Decoding Order Optimization}
 One of
the consequence of the per-user outage and error propagation
effects is that the system goodput cannot scale with SNR due to
potential mutual interference between users. To alleviate this issue,
optimization of transmit power and decoding order in MUD-SIC is
needed. Yet, such optimization problem (taking into account of error
propagation) is extremely complicated and has not been addressed in
the literature.

\item {\bf Macro-Diversity } In multi-cell systems,
macro-diversity (MDiv) enhances signal detection by exploiting the
intercell interference \cite{JR:SHO:96,JR:SHO-analysis:02}. For
instance, packet detection is terminated at each base station
locally and the decoded packets from the base stations (in the
active set) are delivered to a {\em base station controller} where
packet selection is performed. {\em Macro-diversity} is a well
studied technique in CDMA systems with single-user detection at the
base station. However, it is not clear how the MDiv could alleviate
the error propagation effects in the multi-cell network with MUD-SIC.

\end{itemize}

In this paper, we attempt to address the above issues. We
consider an uplink of a multi-cell system with $n_B$ base
stations (each has MUD-SIC) and $K$ mobile users. We derive the
closed-form expressions on the {\em average system goodput} as well as the
{\em per-user packet outage probability} of the MUD-SIC detection
under {\em macro-diversity} and potential error-propagation in the
SIC process. While joint power and decoding order optimization is
a $\cal{NP}$ hard problem, we derive a simple on/off power control
and decoding ordering which is asymptotically optimal w.r.t. the transmit power. Based on
the results, we found that power adaptation, decoding order and
MDiv are important to enhance the system goodput of
MUD-SIC in multi-cell network.

The paper is organized as follows. Section \ref{sect:system-model}
outlines the multi-cell system and the base station MUD-SIC
processing. Section \ref{sect:performance-anaylsis} provides the
analysis of the network goodput of the multi-cell system with
MUD-SIC and MDiv. Section \ref{sect:result-discussion}
presents numerical results on the performance and verify with the
analytical expression. Section \ref{sect:conc} concludes with a
summary of results.

\section{System Model}
\label{sect:system-model}

\subsection{Notation}
 Upper and lower case letters represent random
variables and realizations of the variables, respectively. ${\cal
E}[X]$ denotes the expectation of the random variable $X$. $X_{k:n}$
represents the $k$-th order statistic
($X_{1:n}<X_{2:n},\ldots,<X_{n:n}$) of $n$ ordered random variables. Matrix $\mathbf{\Pi}$ contains vectors $\{\pi_1,\pi_2,\ldots,\pi_b\}$, where $\pi_b$
represents a particular decoding order for base station $b$.
$\pi_b(i)$ gives the user index of users $k$ in the $i$-th decoding iteration at the $b$-th base station and $\pi_b^{-1}(k)$ returns the decoding iteration index of user $k$ at the $b$-th base station.
\begin{figure}[t]\vspace*{-1cm}
  \centering
  \includegraphics[width=3in]{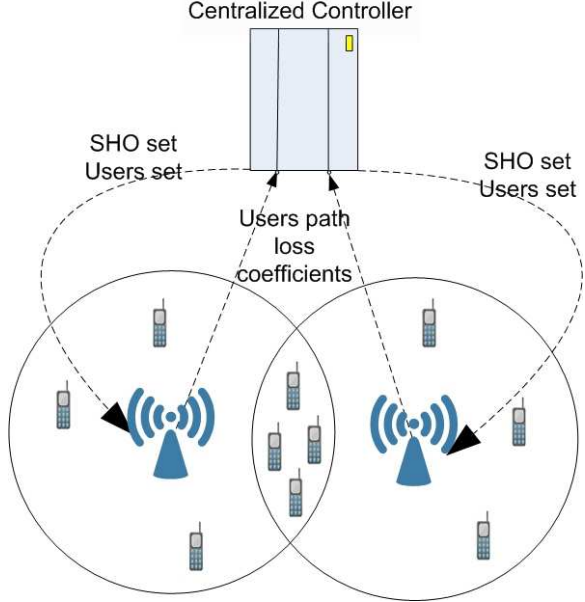}
  \caption{Multi-cell system model with $n_B$ base stations, $K$ mobile users (each has single antenna), and a centralized
controller. }\label{fig:system}\vspace*{-3mm}
\end{figure}

\subsection{Multi-user Multi-cell  Channel Model}
We consider a wireless communication system which consists of $n_B$
base stations, $K$ mobile users, and a centralized controller as
shown in Figure \ref{fig:system}. The base stations and mobile
terminals all have single antenna. The signal received by the $b$-th base station is given by\footnote{
 The proposed system model is a generalized model of CDMA, since the constant spreading factor/ processing gain can be treated as a multiplicative factor and absorbed in the path loss variables. }
\begin{equation}
\label{eqn:receive-signal-model} Y_b=\sum_{i=1}^ {K}
\sqrt{P_{i}g_{i,b}}H_{i,b}X_i+Z_b,
\end{equation}
where $X_i$ is the transmitted signal from the $i$-th mobile
station, $P_i$ is the transmitted power of the $i$-th mobile
station which has range $[0,P_{max}]$, and $Z_b$ is complex
Gaussian noise with zero mean and unit variance at the $b$-th base
station, i.e., $CN(0,1)$. The path loss and shadowing effect, i.e., $g_{i,b}$, between the $b$-th base station
and the $i$-th mobile station can be expressed as
\begin{equation}
\label{eqn:path-loss}
g_{i,b}(\mbox{dB})=\overline{PL_b}(d_0)+10\psi_b\log_{10}\left(\frac{d_i}{d_o}\right)+\omega_\sigma,
\end{equation} where $\overline{PL_b}(d_0)$ is the average path loss at the reference point $d_o$
meters away from the $b$-th base station, $\psi_b$ is the path loss exponent
in the $b$-th cell, $d_i$ is the distance in meters away the
$b$-th BS, and $\omega_\sigma$ denotes the shadowing effect which is modeled as a zero mean Gaussian distributed random variable with standard deviation $\sigma$. In order words, $g_{i,b}$ is log-normal distributed (in dB) with mean $\overline{PL_b}(d_0)+10\psi_b\log_{10}\left(\frac{d_i}{d_o}\right)$ and standard deviation $\sigma$ dB. We model the channel coefficient
$H_{i,b}$ between the $i-$th mobile station and the $b$-th base station as circularly complex Gaussian random variable with zero mean and unit variance.

In general, power and rate adaptation can be performed w.r.t. the product of multipath fading, average path loss and shadowing variables. However, adaptation w.r.t. microscopic fading is challenging especially for fast moving mobiles because the corresponding channel state information need to be updated at the base stations in a frequent manner. These updates increase the signalling overhead significantly and the computational complexity \cite{JR:slow_adaptive,CN:slow_adaptive} at the base stations. As a result, in this paper we assume that the power and data rates of the $K$ users are adaptive w.r.t. long-term fading (path loss and shadowing).


\subsection{Centralized Controller Processing }
The centralized controller is responsible for determining a user
assignment set of each base station\footnote{Mapping of the $K$
users w.r.t. $n_B$ base station is not the focus of this
paper in which users are assumed to be associated with the
strongest base station. For a discussion on mapping algorithm,
please refer to \cite{JR:Vince:camp-on-station}. } and a set of
users who need MDiv to enhance the performance. The $b$-th base station should pass
the estimated macroscopic fading coefficients (average path loss and shadowing) from all
$K$ users to the centralized controller.
{After collecting all the macroscopic fading
information from the $n_B$ base stations, the centralized
controller compares the differences of average path loss and shadowing effect, i.e., $g_{i,b}$, between each mobile
user and all the base stations with a predefine threshold
$\triangle_{\mbox{threshold}}$, and then sends out the MDiv users list to all base stations. Furthermore, for those
mobile users who require MDiv, the decoded messages are passed to
the centralized controller from the corresponding base stations. Then the controller selects a successfully decoded
packet based on the Cyclic Redundancy Check (CRC) field. Since multiple base stations are decoding the same message for a user who demands MDiv and only the correct decoding messages are selected, a form of selection diversity protection is achieved.

\subsection{MUD-SIC Processing and Per-User Packet Error Model}
In this paper, we assume that the base stations are equipped with
synchronous multi-user detector with successive interference
cancellation. Furthermore, we assume that the base stations have
 knowledge of the channel statistic of multipath fading, average path loss and shadowing for all mobile users by long term measurement. On the other hand, the
mobile stations do not have channel state information (CSI) and power allocation in the uplink are calculated at the base station and fed
forward to the mobile stations. The received signal at the $b$-th
base station is given by
\begin{equation}
Y_b=\hspace*{-3mm}\underbrace{\sum_{i\in A_b}
P_{i}g_{i,b}H_{i,b}X_i}_{\mbox{Intra-cell signal}}\hspace*{-3mm}+\hspace*{-3mm}
\underbrace{\sum_{i\notin
A_b}P_{i}g_{i,b}H_{i,b}X_i}_{\hspace*{3mm}\mbox{Inter-cell interference}}\hspace*{-3mm}+
\hspace*{-3mm}\underbrace{Z_b}_{\mbox{Gaussian noise}},
\end{equation}
where $ A_b$ is a user set (including the users which perform
MDiv) that are associated with the $b$-th base station.

The instantaneous channel capacity between the $b$-th base station
and the $k$-th user is given by the maximum mutual
information\footnote{The maximum mutual information can be achieved
if we assume Gaussian random codebook is used.} between the channel
input $X$ and channel output $Y$.
Hence, for a given decoding order
$\pi_b=\{\pi_b(1),\pi_b(2),\ldots,\pi_b(u_b)\} $ and user assignment set $A_b$
with cardinality $u_b$, the instantaneous channel capacity between
the $b$-th base station and the user $j$ in the $\pi_b^{-1}(j)$-th decoding iteration is ${\cal
C}_b(\textbf{H,G},\pi_b,j)=$
\begin{eqnarray}
\label{eqn:channel-capacity} \hspace*{-8mm}&&\log_2\hspace*{-1mm}
\left(\hspace*{-1mm}1\hspace*{-1mm}+\hspace*{-1mm}\frac{P_{j}|H_{\pi_b^{-1}(j),b}|^2g_{\pi_b^{-1}(j),b}}{1+{\widetilde{W_{\pi_b^{-1}(j)}^{\pi_b}}} + {
\Phi}_b(\textbf{H,G},\pi_b,j)+{
\Omega}_b(\textbf{H,G})}\hspace*{-1mm}\right)
\end{eqnarray}
where $\textbf{H}$ is the channel state information at the receiver (CSIR) matrix, $\textbf{G}$ is the average path
loss and shadowing matrix, ${
\Phi}_b(\textbf{H,G},\pi_b,j)={\underset{i=\pi_b^{-1}(j)+1}{\sum_{i\in
A_b}^{u_b}}P_{i}|H_{\pi_b(i),b}|^2g_{\pi_b(i),b}}$ is the \emph{undetected signal}, ${
\Omega}_b(\textbf{H,G})={\sum_{i\notin
A_b}P_{i}g_{i,b}|H_{i,b}|^2}$ is the \emph{inter-cell interference}, and ${\widetilde{W_{\pi_b^{-1}(j)}^{\pi_b}}}$ denotes the accumulated {\em undecodable
interference} after $\pi_b^{-1}(j)-1$ decoding iterations.

 In this paper, we assume packet errors are contributed by channel outage which is a systematic error and
cannot be avoided even when a capacity achieving coding is applied
to protect the packet. As a result, traditional system performance measure
using ergodic capacity may not be a good choice in this situation
since it fails to account for the penalty of packet errors. In
order to model the effect of packet errors, we consider the
performance in terms of the system goodput (bit/s/Hz successfully received).

We model the \emph{undecodable interference} and \emph{per-user goodput} as follows.  The \emph{undecodable interference}
at the $b$-th base station of user $j$ in the $\pi_b^{-1}(j)$-th decoding iteration is
\begin{eqnarray}
{\widetilde{W_{\pi_b^{-1}(j)}^{\pi_b}}}&=&\sum_{i=1}^{\pi_b^{-1}(j)-1}P_{\pi_b^{-1}(i)}|H_{\pi_b^{-1}(i),b}|^2g_{\pi_b^{-1}(i),b}\notag\\
&&\times{\cal I}\left\{r_{\pi_b^{-1}(i)}> {\cal
C}_b(\textbf{H,G},\pi_b,\pi_b^{-1}(i))\right\}.
\end{eqnarray}
For the per-user goodput of user $k$, let $B_k$
denotes the MDiv base station assignment list and the instantaneous goodput
of a packet transmission (bit/s/Hz successfully delivered) to the
$b$-th base station is given by

\begin{equation}
\label{eqn:goodput} \rho_{k}=r_{k} \times\big[1- \prod_{b
\in {{B}}_k}{\cal I}\left\{r_{k}> {\cal
C}_b(\textbf{H,G},\pi_b,k)\right\}\big],
\end{equation}
where $r_k$ is the transmitted data rate of user $k$, which
is a function of the average path loss and shadowing realization only. ${\cal I}\{\cdot\}$ is an
indicator function that evaluates to 1 when the event is true and 0
otherwise. In (\ref{eqn:goodput}), we can see that the goodput of
user $k$ depends on a set of base stations ${ B}_k$\footnote{The
cardinality of ${ B}_k$ is one if user $k$ is assigned to one
base station and no MDiv will be performed.} if the user is
performing MDiv, otherwise the goodput of this user only depends
on one base station. If strong error correction code is applied to
the packet, the conditional average packet error rate (PER) of the
user $k$ (conditioned on the path loss and shadowing realization) can be expressed as
\begin{eqnarray}
\label{eqn:per-outage-approx}
&&\overline{\mbox{PER}_k}(r_k,P_k;\mathbf{G})\approx\overline{ P_{out_k}}(r_k,P_k;\mathbf{G})\nonumber\\
&&\hspace*{-6mm}=\sum_{\pi_b\in {{B}}_k}\prod_{b \in
{{B}}_k}\left\{\Pr\left[r_{k}> {\cal
C}_b(\textbf{H,G},\pi_b,k)|\pi_b,\textbf{G}\right]\Pr(\pi_b)\right\},
\end{eqnarray}
where the first summation accounts for all the possible
combinations of decoding order in $|{ B}_k|$ number of MDiv
stations. Therefore, the average system goodput (conditioned on
the path loss and shadowing matrix $\textbf{G}$) is given by
\begin{eqnarray}
&&U_{gp}(P,R,\mathbf{\Pi};\mathbf{G})={\cal E}_{H} \left[ \sum_{k=1}^K \rho_{k} \vert \textbf{G} \right]\nonumber \\
&=&{\cal E}_{H}\left\{\sum_{k=1}^K
r_k\left({1-\overline{P_{out_k}}(r_k,P_k;
\mathbf{G})}\right)|\textbf{G}\right\} \label{eqn:goodput-define}.
\end{eqnarray}

Note that the average system goodput and PER are both functions of
the transmission power of users and the decoding order. In the
next section, we shall derive the optimal transmit power of each
user and the asymptotically optimal decoding order w.r.t. the transmit power.

\section{Performance Analysis}
\label{sect:performance-anaylsis} In this section, we shall
analyzes the average system goodput and per-user outage
probability of the MUD-SIC system taking into account of
transmission power, potential error propagation and
macro-diversity.
\subsection{Optimal Power Transmission Level with MUD-SIC under Macro-Diversity}
Traditionally, power control is employed to eliminate the near/far
problem by maintaining equal received SINR among all mobile users
when base stations are configured to perform single user detection
\cite{JR:CDMA:91}. On the other hand, for ML detection at the base
station, the optimal power control (under peak power constraint) to
maximize the ergodic sum capacity is simply for each user to
transmit at its maximum power \cite{CN:warrier98capacity}. Yet, in
our case of outage-limited MUD-SIC with potential error-propagation,
it is not obvious if all the users should transmit at their maximum
power due to potential interference in the SIC process. In the
following lemma, we prove that a simple on/off power control is
asymptotically optimal with respect to high transmit power in the outage limited case.
\begin{Lem} [Optimal Power Allocation]
 With the same peak power constraint $0\le P_k\le P_{\max}$ for all users, the optimal power allocation that maximizes the instantaneous mutual information in the outage-limited MUD-SIC system (with potential error propagation) is given by the simple on/off rule:

\begin{equation}
P_{k}=\{0,P_{\max}\},\quad \forall k
\end{equation}
This lemma suggests that a user either transmits at full power or
does not transmit at all. \label{Lem:optimal-power}
\end{Lem}

\begin{proof}
Please refer to Appendix \ref{appen:opt-power}.
\end{proof}

\subsection{ Asymptotically Optimal Decoding Order with MUD-SIC under Macro-diversity}
 In the existing
literature, the decoding order of successive interference
cancellation is usually designed to either minimize the transmit
power subject to performance requirement constraints or to maximize
system capacity with power constraint. In
\cite{JR:joint-power-ordering}, the authors show that solving for
 optimal decoding order is $\cal NP$-hard when the decoding
order is jointly optimized with power allocation, but can be
approximated by means of the discrete stochastic approximation (DSA)
algorithm. In \cite{CN:fast-fading-SIC}, the authors show that for
any point on the boundary of the capacity region, the optimal
decoding policy is successive decoding with the same decoding order
of users for all channel, when the mobile station has perfect CSIT.
 However, these results failed to account for the packet errors in slow fading channels.
Furthermore, due to the mutual coupling of the outage events in
the MUD-SIC processing, the optimal decoding order, which is given
by $\mathbf{\Pi}^*=\arg \underset{\mathbf{\Pi}}{\max}\
U_{goodput}(P,R,\mathbf{\Pi};\mathbf{G})$, is very complicated and requires
exhaustive search in general. Yet, we shall show in Lemma
\ref{Lem:asym-optimal-decode-order} that a simple decoding
ordering would be asymptotically optimal for large transmit power.
\begin{Lem}[Asymptotically Optimal Decoding Order]
For a given path loss realization $\mathbf{G}$, let
$A_b(\mathbf{G}) =\{1,2,\ldots,\mu_b\}$ be the set of active users (users
with non-zero transmit power). Suppose all the users have the same
conditional average PER requirement, i.e.,
$\overline{\mbox{PER}}_k(r_k,P_k; \mathbf{G}) = \epsilon$, then
the following decoding order is asymptotically optimal for
sufficiently large $P_{max}$.

\begin{eqnarray}
\pi^*_b(j)={\arg
\max_{k\in[1,K]\setminus\{\pi_b(1),\pi_b(2),...,\pi_b(j-1)\}}}\quad\gamma_k
\label{eqn:asym-optimal-decoding}
\end{eqnarray}
\label{Lem:asym-optimal-decode-order}
where $\gamma_k = P_{max}|H_{k,b}|^2g_{k,b}$ is the instantaneous
receive SNR of all active users.
\end{Lem}
\begin{proof}
Please refer to Appendix \ref{lem:proof-optimal-decode-order}.
\end{proof}
\begin{figure}[t]
  \centering
  \includegraphics[width=3in]{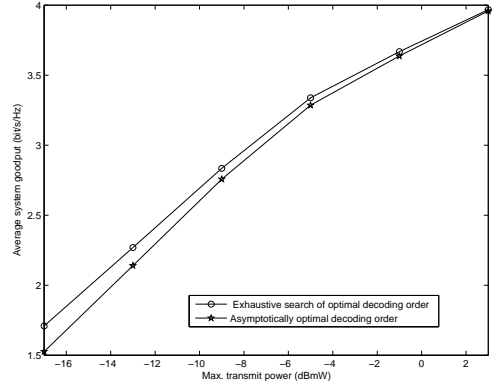}
  \caption{A comparison of asymptotical optimal decoding order and exhaustive search of optimal decoding order in a two cells system. Average system goodput versus max. transmit power with  K=10. Each user is on/off power controlled and with outage requirement 5\%.  }\label{fig:asymptotic-compare}
\end{figure}
While the decoding rule in (\ref{eqn:asym-optimal-decoding}) is only
asymptotically optimal, we show in Figure
\ref{fig:asymptotic-compare} that the decoding rule in
(\ref{eqn:asym-optimal-decoding}) achieves close-to-optimal
performance even in moderate SNR.

In order to characterize the per-user outage probability, let's
define $S_i=\{0,1\}$ as the $i$-th stage iteration decoding event
with $S_i=1$ denotes successful decoding and $S_i=0$ denotes
decoding failure. Given the asymptotically optimal decoding order policy in
(\ref{eqn:asym-optimal-decoding}), we assume that for user
$\pi^*_b(i)$ fails in the $i$-th decoding iteration, then we can
declare packet error for all the remaining users in the same base
station. This assumption cause a neglectable sub-optimality to the
system performance which can be verified by numerical simulation,
however, it can provide a tractable analysis expression and provide
some important insights regarding the system performance.

Next, we define the event $S_i$ which is given by
\begin{equation}
S_i = {\cal I}\left\{ r_{\pi^*_b(i)}<\log_2\left( 1+
\mbox{SINR}_{\pi^*_b(i)}\right)\right\}=\{0,1\},
\label{eqn:event1}
\end{equation}
where $\mbox{SINR}_{\pi^*_b(i)}=\frac{\gamma_{\pi^*_b(i)}}{1 + \sum_{j<i}\gamma_{\pi^*_b(j)}(1-S_j)
+ \sum_{j>i}\gamma_{\pi^*_b(j)}} $ is the signal-to-interference plus noise ratio and ${\cal I}(\cdot)$ is the {\em indicator function} which is 1 when
the event is true and 0 otherwise. Since we assume that any packet error before the $i$-th stage will result decoding error in the remaining stages, we can define the following event
\begin{equation}
{\cal O}_i = {\cal I}\left\{r_{\pi^*_b(i)} < \log_2\left(1 +
\frac{\gamma_{\pi^*_b(i)}}{1 + \sum_{j>i}\gamma_{\pi^*_b(j)}}
\right)\right\}. \label{eqn:event3}
\end{equation}
 Based on the above assumption and event definition, we can duce that:
\begin{equation}
S_i=0 \Rightarrow {\cal O}_1\cup {\cal O}_2\cup....\cup {\cal O}_i.
\label{eqn:event2}
\end{equation}
Therefore, the packet outage probability of user $k$ is given by
\begin{eqnarray} &&\overline{P_{out_k}} (r_k,P,k;\mathbf{G})\nonumber\\
&&\hspace*{-3mm}\sum\limits_{\pi_b^* \in { B}_k}\prod_{b\in { B}_k}
{\sum\limits_{i=1}^{\pi_b^{*-1}(k)} {\Pr \left[ {{\cal O}_1\cup
{\cal O}_2\cup....\cup {\cal O}_i =0\vert \pi_b^* } \right]\Pr
(\pi_b^* )} }\nonumber\\&&\hspace*{-3mm}\le \sum\limits_{\pi_b^*\in
{ B}_k }\prod_{b\in { B}_k} {\sum\limits_{i=1}^{\pi_b
^{*-1}(k)} {\sum\limits_{j=1}^i {\Pr \left[ {{\cal O}_j =0\vert
\pi_b^* } \right]} \Pr (\pi_b^* )} }\label{eqn:per_user_outage}
\end{eqnarray}
By substituting (\ref{eqn:per_user_outage}) into (\ref{eqn:goodput-define}), the average system goodput under the asymptotically
optimal decoding order is given by
\begin{eqnarray}
\label{eqn:average-system goodput} &\hspace*{-10mm}&U_{gp}(P,R,\mathbf{\Pi}^*;\mathbf{G})
=\sum\limits_{k=1}^{K} {r_k (1-\overline{P_{out_k}}
(r_k,P,k;\mathbf{G})} )
\\
\hspace*{-10mm}&\ge&\hspace*{-3mm} \sum_{k=1}^{K}r_k\big(1\hspace*{-1mm}-\hspace*{-1mm}\sum\limits_{\pi_b^*\in { B}_k }\prod_{b\in { B}_k} {\sum\limits_{i=1}^{\pi_b ^{*-1}(k)}
{\sum\limits_{j=1}^i {\Pr \left[ {{\cal O}_j =0\vert \pi_b^* }
\right]} \Pr (\pi_b^* )}
}\big).\nonumber
\end{eqnarray}

 \subsection{Per-user
PER and Average System Goodput  }

Under the asymptotically optimal decoding order in Lemma
\ref{Lem:asym-optimal-decode-order}, the average system goodput and
per-user outage probability can be expressed in term of the
conditional outage probability. In order to solve the per-user
outage probability and average system goodput, we should obtain the
closed form expression of the conditional outage probability. For a
given asymptotically optimal decoding order $\pi^*_b$, the
conditional outage probability of user $k$ in the $j$-th iteration
can be expressed as:
\begin{eqnarray}
\label{eqn:outage_conditional}
&& \hspace*{-3mm}\Pr \left[ {{\cal O}_j =0\vert \pi^*_b } \right]=\Pr (r_k >C_{\pi_b^*(k)} ({\rm
{\rm{ \bf H}}},{\rm {\bf G}},\pi^*_b,k)\vert \pi^*_b
)\nonumber\\
&&\hspace*{-3mm}=\Pr\left \{
\gamma_{\pi^*_b(j)}-\vartheta_{\pi^*_b(j)}\sum_{l=j+1}^{u_b}\gamma_{\pi^*_b(l)}<\vartheta_{k}\right\}
 \end{eqnarray} where $\vartheta_{k}=2^{r_k}-1$. In general, the conditional outage probability involve $\mu_b$
dimensions nested integration which is complicated and non-traceable
when the dimension of integration grows. However, by taking the
advantage of the additive Markov chain property from the exponential
random variable order statistics, the conditional outage probability
can be calculated by a one dimensional integration. We first
introduce the following lemma.
\begin{Lem}
[ Closed-form Expression of Conditional PER]
\label{Lem:transformation}

The conditional outage probability of user $k$ in the $j$-th iteration in (\ref{eqn:outage_conditional}) can be written in a summation of
exponential functions which is given by
\begin{eqnarray}
\Pr \left[ {{\cal O}_j =0\vert \pi^*_b}
\right]&=&1-\sum_{l=j,\upsilon_l>0}^{\mu_b}\Psi_l\frac{\beta_l}{\upsilon_l}
\exp(-\frac{ \vartheta_{k}\beta_l}{\upsilon_l})
\end{eqnarray}
where $\Psi_l=\prod_{i=j,i\ne
l}^{\mu_b}\frac{\upsilon_l}{\upsilon_l-\frac{\beta_l}{\beta_i}\upsilon_i},\upsilon_l=\frac{1-l\times
\vartheta_{k}+j\times\vartheta_{k}}{l}$,
$\beta_l=\frac{\sum\limits_{u=1}^l {\frac{1}{g_{\pi^*(u),b}}}}
{P_{\max}l}$, and $\vartheta_{k}=2^{r_k}-1$.
\end{Lem}
\begin{proof}
Please refer to Appendix
\ref{appendix:pf-lem:transformation-order-statistic}.
\end{proof}

 After obtaining the
closed-form of the conditional outage probability, we need to
calculate the probability of a particular decoding order which is
summarized in the following:
\begin{Lem}[ Probability of a Decoding Order Policy $\pi_b$ ]
\label{Lem:decoding-order-calculation} Consider a set of independent non-identical distributed (i.ni.d.)
exponential random variables ${X_1,X_2,X_3,\ldots,X_{\mu_b}}$ which has a
p.d.f. as defined in (\ref{eqn:i.ni.d-exp}). From \cite{book:order-book}, the
probability of a particular order
$X_{i_1:\mu_b}<X_{i_2:\mu_b}<...<X_{i_{\mu_b}:\mu_b}$ is given by
\begin{eqnarray}
&&\Pr
(X_{i_1:\mu_b}<X_{i_2:\mu_b}<\ldots<X_{i_{\mu_b}:\mu_b})\\
&=&\hspace*{-2mm}\frac{\beta_{i_1}\beta_{i_2}\beta_{i_3}\ldots\beta_{i_{\mu_b}}}{(\beta_{i_1}+\beta_{i_2}+\ldots+\beta_{i_{\mu_b}})(\beta_{i_2}+\beta_{i_3}+\ldots+\beta_{i_{\mu_b}})\ldots\beta_{i_{\mu_b}}}\notag
\label{eqn:ordering-prob}
\end{eqnarray}
\end{Lem}
As a result, the per-user outage probability is given by the
following lemma.

\begin{Lem}[Per-User Conditional PER with Macro-diversity]
The average packet error probability of user $k$ under the
asymptotically optimal decoding order policy ${\mathbf\Pi}^*$ is given
by:
\begin{eqnarray}
\label{eqn:outage-prob}
 &&\hspace*{-7mm}\overline{P_{out_k}} (r_k ,P_{max};\mathbf{G}) \notag\\
 &&\hspace*{-7mm}\le\hspace*{-3mm} \sum\limits_{\pi_b^*\in { B}_k }\prod_{b\in { B}_k} {\sum\limits_{i=1}^{\pi_b{^*} ^{-1}(k)}
{\sum\limits_{j=1}^i {\Pr \left[ {{\cal O}_j =0\vert \pi_b^* }
\right]} \Pr (\pi_b^* )} }\nonumber\\
&&\hspace*{-7mm}=\hspace*{-3mm} \sum\limits_{\pi_b^*  \in{ B}_k}\prod_{b\in { B}_k} {\sum\limits_{i=1}^{\pi_b^{*{-1}}(k)} {\sum\limits_{j=1}^i
\left\{1-\sum_{l=j,\upsilon_l>0}^{\mu_b}\Psi_l\frac{\beta_l}{\upsilon_l}
\exp(-\frac{ \vartheta_{k}\beta_l}{\upsilon_l}) \right\} \Pr
(\pi_b^*
 )} }
\nonumber \\
\end{eqnarray}
\end{Lem}
where $\Pr(\pi_b ^\ast) $ is given in equation
(\ref{eqn:ordering-prob}) and $\vartheta_k=2^{r_k}-1$.
Therefore, the average system goodput
can be summarized by the following theorem:
 \begin{Thm}[Lower Bound for the Average System Goodput ]
\begin{eqnarray}
\label{eqn:average-system goodput}
&&\hspace*{-8mm}U_{gp}(P,R,\mathbf{\Pi}^*;\mathbf{G}) =\sum\limits_{k=1}^{K} {r_k
(1-\overline{P_{out}} (r_k,P_k;\mathbf{G})} )
\nonumber\\
 \hspace*{-8mm}&\ge&\hspace*{-3mm} \sum_{k=1}^{K}r_k\Bigg\{1-\sum\limits_{\pi_b ^* \in{ B}_k } \prod_{b\in { B}_k}{\sum\limits_{i=1}^{\pi_b^{*{-1}}(k)} {\sum\limits_{j=1}^i {\Pr
\left[ {{\cal O}_j =0\vert \pi_b^* } \right]} \Pr (\pi_b^* )}
}\Bigg\}\nonumber\\
 &=&\hspace*{-3mm} \sum\limits_{k=1}^{K} r_k\times\left\{\rule{0cm}{18pt}1-\right.\notag\\
 &&\hspace*{-8mm}\bigg.\sum\limits_{\pi_b ^* \in{ B}_k}{\underbrace{\prod_{b\in { B}_k}{\sum\limits_{i=1}^{\pi_b^{*{-1}}(k)} {\sum\limits_{j=1}^i
{\sum_{l=j,\upsilon_l>0}^{\mu_b}\Psi_l\frac{\beta_l}{\upsilon_l}
\exp(-\frac{ \vartheta_{k}\beta_l}{\upsilon_l}) } } }
}_{\mbox{Selection diversity protection}}\Pr
(\pi_b^* )}\rule{0cm}{18pt}\bigg\}
\end{eqnarray}
 \end{Thm}
From the above expression, the second summation represents the
system goodput corresponds to each decoding permutation of the
decoding rule in equation (\ref{eqn:asym-optimal-decoding}). The
product term in (\ref{eqn:average-system goodput}) offers MDiv
protection as a packet has to fail in all the base stations to declare packet
error.

\begin{Remark}

The data rate $r_k(\mathbf{G})$ can be determined by solving the per-user
conditional packet error requirement
$\overline{P_{out_k}}(r_k,P_{max};\mathbf{G})=\epsilon$.
 \end{Remark}

\section{Results and Discussions}
\label{sect:result-discussion}

In this section, we evaluate the theoretical results in the
preceding section using simulations. We consider a multi-cell system with 2 base stations. Every cell
has radius of 1 km and path loss exponent 3.6. Assume that the
minimum distance from a mobile station to the home base station is
30 m,
 the average path loss of a particular user in the cell has a dynamic
range up from -48 dB to -103 dB. The noise power level is equal to -105 dBm. The log-normal shadowing is assumed to have a standard derivation 8 dB. There are $K$ active users
uniformly distributed in the cells and the distance the mobile and
$b$-th base station and the $k$-th mobile user is $d_{k,b}$. All the channel
fading coefficients $\{H_1,H_2,...,H_K\}$ are generated as i.i.d.
complex Gaussian random realizations with zero mean and unit
variance. Average system goodput is obtained by counting the
number of packets which are successfully decoded by the base
station for all users and average the result over both macroscopic
and microscopic fading. In the simulation, each point is obtained
by averaging 100000 macroscopic and microscopic realizations.

\begin{figure}[t]
  \centering
  \includegraphics[width=3in]{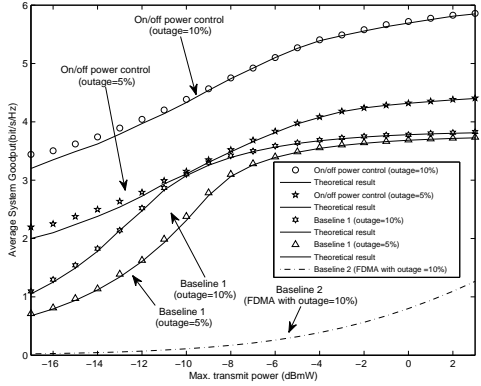}
  \caption{A comparison of on off power control and baseline 1 power control
  scheme. Baseline 1 represents traditional CDMA power control algorithm in which the
transmit powers of all active users are adjusted such that the
received SINR of them are the same at base station.
{Baseline 2 represents FDMA system and each user transmit at its peak power.}  Average system
goodput versus max transmit power in a two cells system with K=10,
outage requirement 5\% or 10\% for
$\triangle_{\mbox{threshold}}=\infty$ (no
MDiv).}\label{fig:equal_rate_multi_rate}
\end{figure}

\subsection{ Average System Goodput}
Figure \ref{fig:equal_rate_multi_rate} illustrates the average
system goodput versus the  transmit power (dBmW) of mobile user for
$K=10$ with asymptotical optimal decoding order. Each curve in the
graph represents different type of power control with same target
outage probability for all user (5\% or 10\%). The optimal data rate of each user is obtained by numerical method such as Newton method in solving equation (\ref{eqn:average-system goodput}) for $\overline{P_{out_k}}(r_k,P_{max};\mathbf{G})=\epsilon, \ \forall k$. We compare the
performance of the proposed design with a conventional baseline 1 CDMA
power control algorithm\footnote{
 The data rate in the simulation of CDMA is set to a value such that the outage requirement can be fulfilled for the weakest user. For the spreading in the CDMA system, we assume the synchronized orthogonal spreading codes are used and the spreading factor is always equal to the number of users. Therefore, the orthogonal multiple access incurs no loss in total system capacity for equal rate and equal SINR users \cite{JR:CDMA}. } in which the transmit powers of all users
are adjusted such that the received SINR of them are the same at
base station. For the baseline 1, the system goodput grows with SNR at
small SNR\footnote{Because all active users are transmitting at
their max power, therefore the SNR of each user is directly
proportional to the max power. } but quickly saturated at moderate
SNR. This is because the performance is always limited by the
weakest users. On the other hand, the goodput performance of the
proposed on/off power control scheme does not saturate even at high
SNR regime. It can be explained that in the proposed on/off power
control, strong users do not required to decrease the transmission
power to maintain the same SINR as those weak users, this factor
contribute significantly to the system goodput.
Furthermore, we compare the proposed design with a baseline 2 (FDMA system) where each user transmits at its peak power. Although multiple access interference does not exist in the FDMA system due to orthogonal transmission, it has a very low spectral efficiency. On the contrary, the proposed design provides a substantial performance gain compared with the FDMA system in the interference limited environment.

\begin{figure}[t]
  \centering
  \includegraphics[width=3in]{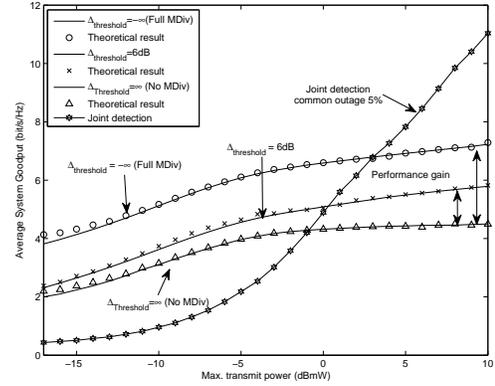}
  \caption{Average system goodput versus max transmit power with different MDiv
threshold in a two cells system, K=10. Each user has a outage
requirement $=5\%$ and is power controlled by on/off transmission. The
double sided arrow represents the performance gain due to MDiv.
Common outage is declared when the rate vector is outside the
instantaneous capacity region.
}\label{fig:goodput_SNR-diff-handoffthreshold}
\end{figure}
Figure \ref{fig:goodput_SNR-diff-handoffthreshold} shows the average
system goodput versus the transmit power with different MDiv
threshold ($\triangle_{\mbox{threshold}}$). Each user is power
controlled by the on/off scheme and there is 5\% outage probability
requirement. We compare the performance of the proposed design with
a system that does not perform MDiv in which all the inter-cell
users are treated as interference. For the system without MDiv, the
average system goodput saturated at high SNR because strong
interference from inter-cell becomes a dominate factor in the system
performance. On the contrary, the average system goodput of the
proposed design increase with the transmit power when MDiv is
performed in the base station. The reason is that strong
interference is regarded as desired user signal and it will be
decoded by corresponding base station. Furthermore, the optimal
power control (either full power transmission or completely silent)
create a high disparities\footnote{A high disparities received power
can significantly increase the system capacity for MUD-SIC receiver
\cite{CN:warrier98capacity}.} received power at the base station and
strong enough interference environment for MDiv to exploit.
Therefore, the system goodput has a significant gain when MDiv is
performed in multi-cell environment. Furthermore, the goodput of the
joint ML detection (which consider common outage\footnote{Common
outage is declared as rate vector is outside the instantaneous
capacity region.}) is plotted for comparison. In low SNR regime, the
SIC outperforms the joint ML detection. This is because in the joint
ML detection, a common outage will be declared if the rate vector
lies outside the instantaneous capacity region. Hence, the outage
performance of the joint ML detection is always limited by the
weakest users. On the other hand, the SIC approach consider per-user
outage and packets for some users may be decoded correctly even
though the rate vector lies outside the capacity region. In high SNR
region, the performance of SIC is limited by strong interference
from both intra-cell and inter-cell interference. Nevertheless,
using MDiv, the performance of the SIC scheme can be improved at
high SNR regime. On the other hand, the joint ML detection does not
suffer from multi-user interference and hence the performance is able to scale with SNR.
\begin{figure}[t]
  \centering
  \includegraphics[width=3in]{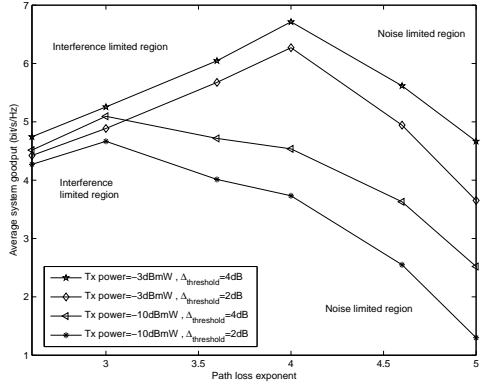}
  \caption{Average system goodput versus path loss exponent in a two cells system and K=10. Each user has a outage requirement 10\%. Transmit power of
users are on/off power controlled and the max power are fixed at
-3 dBmW and -10 dBmW. respectively.
}\label{fig:goodput_path-loss}
\end{figure}

\begin{figure}[t]
  \centering
  \includegraphics[width=3in]{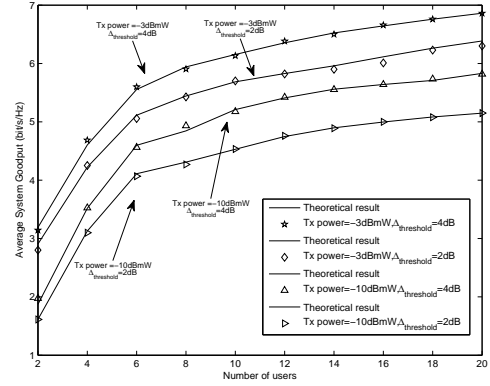}
  \caption{Average system goodput versus number of users in a two cells system. Each user has a outage requirement 10\%. Transmit power
of users are on/off power controlled and the max power is fixed at
-3 dBmW and -10 dBmW, respectively. }\label{fig:goodput-user}
\end{figure}
Similarly, Figure \ref{fig:goodput_path-loss} shows the average
system goodput versus different value of path loss exponent for
$K$=10. Along all the curves, the same user transmit power is fixed
at -10 dBmW and -3 dBmW respectively. It is very interesting that the
average system goodput first increase with the value of path loss
exponent and then decrease when the path loss exponent is beyond
certain value. This counter intuitive result is due to the fact that
when the path loss exponent increases, both desired signal and
interference signal received by base station decreases. However, the
attenuation of interference occurs to be larger than desired signal
because interference users are usually located far away from desired
home base station. As path loss exponent increases, the operating
region of the system is shifting from interference limited region to
noise limited region, and the desired users signal strength
attenuate to a level that high data rate communication is
impossible, and it results in a decreasing trend of average system
goodput.

Figure \ref{fig:goodput-user} depicts that average number of
system goodput versus the number users in a two cells system.
Similarly, the transmit power of an active user is again fixed at
-10dBmW and -3dBmW respectively. It can be observed that the
system goodput gain due to MDiv is not significant when the number
of user is small, especially users are transmitting at low power
(-10dBmW). When the transmit power is low, signal strength of
interference can not satisfy the MDiv threshold requirement, so
there is nearly no MDiv performed in the base stations. However,
when the number of users increases, it is more likely that there
exists a user who locates near the cell boundary, creates large
interference to neighboring cells. Therefore, base stations can
take advantage of the strong interference and perform MDiv to
improve the system goodput. On the other hand, there is a
diminishing return in the system goodput when the number of users
increases, particular in high transmit power with small MDiv
threshold value (2dB). This is due to the fact that base stations
do not fully utilize the benefit of strong interference by setting
a small threshold value\footnote{A small threshold value implies a
few users are satisfied with the MDiv requirement. }, therefore
interference can not be decoded and causes the degradation in
system performance.

\section{Conclusion}
\label{sect:conc} In this paper, a generic multi-cell system with $K$
client users, $n_B$ base stations and a centralized controller is
considered. Based on the asymptotic optimal decoding order with respect to the transmit power, we
incorporate the mathematical tool of order statistics to obtain the
closed-form solution of system performance. Numerical simulations
result are obtained to verify the analytical expressions.  The
closed form solutions allow efficient numerical evaluations to find
out how the system performance is affected by the system parameter
such as number of users and path loss exponent. From the results, we
see that in interference limited region (users transmit at high
power), MDiv improves the system goodput significantly by introducing
macro-diversity protection to alleviate the consequences of error
propagation. Furthermore, system with MDiv allows more users to be
served at the same time through taking advantage of strong
interference.

\appendix
\subsection{ Proof of Lemma \ref{Lem:optimal-power}
}  \label{appen:opt-power} Note that since our power constraint is
instantaneous, average system goodput maximization is the same as
maximize the instantaneous goodput for each fade
vector. It can be observed in equation (\ref{eqn:goodput}) that the system goodput is contributed by the instantaneous channel capacity and transmitted data rate. In fact, the system goodput is upper bounded by the instantaneous channel capacity. In the following, we would like to find the optimal power allocation which can maximize the upper bound of the average system goodput.

For any decoding order $\pi_b$ in base station $b$, the
instantaneous mutual information of user $k$ is given by equation (\ref{eqn:channel-capacity}).
Therefore, in high SNR the total instantaneous capacity ${\cal Q}=\sum_{k=1}^{K}\sum_{b\in B_k^*}{\cal
C}_b(\textbf{H,G},\pi_b,k)\approx$
\begin{eqnarray}
&&\hspace*{-10mm}\sum_{k=1}^{K}\sum_{b\in B_k^*}\log_2\hspace*{-2mm}
\left(\frac{P_{k}|H_{\pi_b^{-1}(k),b}|^2g_{\pi_b^{-1}(k),b}}{{\widetilde{W_{\pi_b^{-1}(k)}^{\pi_b}}} + {
\Phi}_b(\textbf{H,G},\pi_b,j)+{
\Omega}_b(\textbf{H,G})}\hspace*{-2mm}\right)
\end{eqnarray}
where $B_k^*$ denotes the base station which has the maximum mutual information\footnote{Since the packet selection is performed when users are involved in the MDiv and a packet can be possibly decoded by more than one base stations, we can focus on the base station who gives the maximum mutual information for user $k$.   } for user $k$. To find the optimal power allocation that maximizes the instantaneous mutual information, we consider the following optimization
problem.
\begin{eqnarray}
\label{eq:original-power-problem}
P^*=\{P_1^*,P_2^*,\ldots,P_K^*\}=\arg
\max_{\{P_1,P_2,...,P_K\}}{\cal Q}
\end{eqnarray}

Differentiating the system capacity twice
with respect to $P_{j}$, which yields $\frac{\partial^2 {\cal Q}}{{\partial
P_j}^2}=\frac{-1}{P_j^2 \ln 2}$
\begin{eqnarray} & +& \hspace*{-3mm}\frac{1}{\ln 2}\sum_{k\neq j}^K\sum_{b\in B_k^*}\frac{(|H_{\pi_b^{-1}(j),b}|^2g_{\pi_b^{-1}(j),b})^2}{(P_j+\sum_{i\in K-\{k\}}^KP_i|H_{\pi_b^{-1}(i),b}|^2g_{\pi_b^{-1}(i),b})^2}\notag\\
&\approx& \hspace*{-3mm}\frac{-1}{P_j^2 \ln 2}\le0. \label{eqn:power_second_derivative}
\end{eqnarray}
It can be observed that the first term in the derivative is the dominating term since the other terms converge to zero much faster with respect to the transmit power of all users in high transmit power regime. Therefore, $\frac{\partial^2 {\cal
Q}}{{\partial P_j}^2}$ is non-positive and the system goodput is a concave function of $P_j$. Similarly, by differentiating the system capacity once with respect to $P_{j}$, we obtain $\frac{\partial {\cal Q}}{{\partial
P_j}}=\frac{1}{P_j \ln 2} $
\begin{eqnarray} &&\hspace*{-3mm}-\frac{1}{\ln 2}\sum_{k\neq j}^K\sum_{b\in B_k^*}\frac{|H_{\pi_b^{-1}(j),b}|^2g_{\pi_b^{-1}(j),b}}{(P_j+\sum_{i\in K-\{k\}}^KP_i|H_{\pi_b^{-1}(i),b}|^2g_{\pi_b^{-1}(i),b})}\notag\\
&\approx& \frac{1}{P_j \ln 2}\ge0. \label{eqn:power_first_derivative}
\end{eqnarray}
As the first derivative is approximately non-negative and the objective function is a concave function with respect to $P_j$, we conclude that   $P_j=P_{max}$ achieves the maximum system capacity.
\subsection{ Proof of Lemma
\ref{Lem:asym-optimal-decode-order}}
\label{lem:proof-optimal-decode-order}
From equations
(\ref{eqn:channel-capacity}), (\ref{eqn:per-outage-approx}) and
(\ref{eqn:goodput-define}), it can be observed that the optimal
decoding order in maximizing the system capacity is equivalent to a
decoding order, which maximize the instantaneous mutual
information in each decoding iteration at each base station. The
total mutual information in the $b$-th station can be expressed
as:
\begin{eqnarray}
\sum_{k=1}^{K}\sum_{b\in B_k^*}{\cal C}_b(\textbf{H,G},\pi_b,k)\nonumber
\end{eqnarray}
At asymptotically high transmit power ($P_{max}\rightarrow \infty $), 
given a particular decoding order $\pi_b$ and the
\emph{accumulated undecodable}\footnote{Undecodable interference
is due to cancellation error in the previous decoding stage.}
interference $U_1^{\pi_b}$ in the first iteration, the
channel capacity of the user in the first decoding
iteration at the $b$-th base station is given by:
\begin{eqnarray}
\label{eqn:asymp:capacity} && \hspace*{-8mm}{\cal
C}_b(\textbf{H,G},\pi_b,\pi_b(1))\nonumber\\
\nonumber&&\hspace*{-8mm}\approx\hspace*{-1mm}\log_2
\left(1+\frac{P_{max}g_{\pi_b(1),b}|H_{\pi_b(1),b}|^2}{P_{max}\left(U_1^{\pi_b}+
\widetilde{
\Phi}_b(\textbf{H,G},\pi_b,\pi_b(1))+\widetilde{\Omega}_b(\textbf{H,G})\right)}\right)\nonumber\\
&&\hspace*{-8mm}=\log_2
\left(1+\frac{g_{\pi_b(1),b}|H_{\pi_b(1),b}|^2}{U_1^{\pi_b}+
\widetilde{
\Phi}_b(\textbf{H,G},\pi_b,j)+\widetilde{\Omega}_b(\textbf{H,G})}\right)
\end{eqnarray}
where $U_1^{\pi_b}=0$ in the first iteration, $\widetilde{
\Phi}_b(\textbf{H,G},\pi_b,\pi_b(1))={
\Phi}_b(\textbf{H,G},\pi_b,\pi_b(1))/P_{max}$, and $\widetilde{\Omega}_b(\textbf{H,G})={\Omega}_b(\textbf{H,G})/P_{max}$   In order to maximize
the channel capacity of the first iteration, it is equivalent to
select a user to decode according to the following rule:
\begin{eqnarray}
{\pi_b}^*(1)=\arg \underset{k\in[1,\mu_b]} {\max}g_{k,b}|H_{k,b}|^2
\end{eqnarray}
Considers the second iteration of the decoding process. The
\emph{accumulated undecodable interference} has value
$U_2^{\pi_b}\in\{0\quad
 g_{\pi_b(1),b}|H_{\pi_b(1),b}|^2\} $. Therefore, the mutual
 information in the second iteration is given by ${\cal
C}_b(\textbf{H,G},\pi_b,\pi_b(2))=$
\begin{eqnarray}
 &&\hspace*{-12mm}\log_2
\left(1+\frac{g_{\pi_b(2),b}|H_{\pi_b(2),b}|^2}{U_2^{\pi_b}+
\widetilde{
\Phi}_b(\textbf{H,G},\pi_b,\pi_b(2))+\widetilde{\Omega}_b(\textbf{H,G})}\right).
\end{eqnarray}

Similarly, the choice of $\pi_b(2)$ that maximizes the mutual
information is given by:
\begin{eqnarray}
{\pi_b}^*(2)=\arg \underset{k\in[1,\mu_b]\backslash \{\pi_b^*(1)\}}
{\max}g_{k,b}|H_{k,b}|^2
\end{eqnarray}
As such, by induction, the asymptotically optimal decoding
order\footnote{Given the path loss and CSIR realization, the
optimal decoding order should gives the largest number of
successfully decoded users, or equivalently the lowest potentially
accumulated undecoded interference.} is to decode the users
sequentially in decreasing receive SNR as in
(\ref{eqn:asym-optimal-decoding}).

\subsection{Proof of Lemma
\ref{Lem:transformation}
\label{appendix:pf-lem:transformation-order-statistic}
 }
By \cite{book:order-book} and \cite{JR:order_statistics}, for a
given ordered $\Gamma_{1:\mu_b} <\Gamma_{2:\mu_b} <\ldots<\Gamma_{\mu_b:\mu_b}$ channel
gains where $\Gamma_{i}=|H_{i,b}|^2g_{i,b}$, define a new set of
random variables \{$D_1,D_2,...,D_{\mu_b}$ \} to denote the spacing
between $\Gamma_{l:\mu_b} $ and $\Gamma_{l-1:\mu_b}$ as follows:
\begin{eqnarray} \left\{ {\begin{array}{l}
 D_1 =\Gamma_{\mu_b:\mu_b} \\
 D_l =\Gamma_{\mu_b-l+1:\mu_b} -\Gamma_{\mu_b-l:\mu_b,\quad l \mbox{=2,\ldots,$\mu_b$-1}} \\
 \end{array}} \right.
\end{eqnarray}
Then, a linear combination of the spacing is defined as:
\begin{equation}
\label{eqn:spacing-transform} {\cal M}_i =i\{D_i \}
 \end{equation}
where
 $\{{\cal M} \}$ is a set of independent exponential random variables with
p.d.f. given by: \begin{equation} \label{eqn:i.ni.d-exp}
 f_{m_i } (m)=\beta _i \exp(-m\beta _i) ,\qquad \forall m,\beta_i\ge0
\end{equation}
where $\beta_i $ is defined as :
\begin{equation}
\beta_i=\frac{\sum\limits_{u=1}^i {\frac{1}{g_{\pi_b(u),b}}}}
{P_{\max}i}
\end{equation}
Hence, the conditional outage probability can be written as:
\begin{eqnarray}
\label{eqn:conditional-outage -M}
 &&\Pr \left[ {{\cal O}_j =0\vert \pi_b} \right]
=\Pr\left \{ \sum_{l=j}^{\mu_b}{\cal M}_l\upsilon_l<\vartheta_{\pi_b(j)}\vert \pi
_b \right\}\notag\\
&&=\Pr\left \{ {\cal W}_l<\vartheta_{\pi_b(j)}\vert \pi _k
 \right\}
\end{eqnarray}
where $\upsilon_l=\frac{1-l\times
\vartheta_{\pi_b(j)}+j\times\vartheta_{\pi_b(j)}}{l}$, $ {\cal
W}_l=\sum_{l=j}^{\mu_b}{\cal M}_l\upsilon_l$, and $\vartheta_{\pi_b(j)}=2^{r_{\pi_b(j)}}-1$.

Representing $\phi_l(\omega)=\frac{\beta_l}{\beta_l+\upsilon_l
j\omega}$  as characteristic function of ${\cal M}_l\upsilon_l$ ,
then the p.d.f of the ${\cal W}_l$ is given by the inverse Laplace
transform of the following:
\begin{eqnarray} \label{eqn:nested-transform1} f_{{\cal
W}_l}(x)={\cal L}^{-1}\left\{
 \prod_{l=j}^{\mu_b}  \phi_j(\omega)\right\}
\end{eqnarray}
By using the partial-fraction decomposition technique
\cite{book:partial-fraction},
 the conditional outage probability results in a summation of
exponential function which is given by:
\begin{eqnarray}
\Pr \left[ {{\cal O}_j =0\vert \pi _b} \right]&=&\Pr\left
\{{\cal W}_l<\vartheta_{\pi_b(j)}\vert \pi _b  \right\}\nonumber\\
&=&\int_{-\infty}^{\vartheta_{\pi_b(j)}} f_{{\cal W}_l}(x)dx \nonumber\\
&=&\hspace*{-2mm}1-\hspace*{-2mm}\sum_{l=j,\upsilon_l>0}^{\mu_b}\Psi_l\frac{\beta_l}{\upsilon_l}
\exp(-\frac{ \vartheta_{\pi_b(j)}\beta_l}{\upsilon_l})
\end{eqnarray}
where $\Psi_l=\prod_{i=j,i\ne
l}^{\mu_b}\frac{\upsilon_l}{\upsilon_l-\frac{\beta_l}{\beta_i}\upsilon_i}$
are the partial fraction coefficients.
\vspace*{-.6cm}
\bibliographystyle{IEEEtran}
\bibliography{IEEEabrv,multi-cell-SIC-v9}
%
%
\vspace*{-1cm}

\begin{IEEEbiographynophoto}{Derrick Wing Kwan Ng (S'06)  }
received the bachelor degree with First
class honor and Master of Philosophy (MPhil) degree in electronic engineering from the
Hong Kong University of Science and Technology (HKUST) in 2006 and 2008, respectively. He is currently working toward the Ph.D.
degree in the University of British Columbia (UBC). His research interests include cross-layer optimization for wireless communication systems, resource allocation in MIMO and OFDMA wireless system and communication theory.
He received the Best Paper Award at the IEEE Third
International Conference on Communications and Networking in China 2008.  He
was also the recipient of the 2009 Four Year Doctoral Fellowship from the UBC, Sumida \& Ichiro Yawata Foundation Scholarship in 2008 and  R\&D Excellence scholarship from the Center for Wireless Information Technology in the HKUST in 2006.
\end{IEEEbiographynophoto}

\begin{IEEEbiographynophoto}{Vincent K.N.Lau(M'98-SM'01)  }

obtained a B.Eng (Distinction 1st Hons) from the University of
Hong Kong (1989-1992) and a Ph.D. from Cambridge University
(1995-1997). He was with HK Telecom (PCCW) as system engineer from
1992-1995 and Bell Labs - Lucent Technologies as a member of the
technical staff from 1997-2003. He joined the Department of ECE,
Hong Kong University of Science and Technology (HKUST) as an
Associate Professor. At the same time, he is a technology advisor
of HKASTRI, leading the Advanced Technology Team on Wireless
Access Systems. His current research focus is on the robust cross
layer scheduling for MIMO/OFDM wireless systems with imperfect
channel state information, communication theory with limited
feedback as well as cross layer scheduling for users with
heterogeneous delay requirements.
\end{IEEEbiographynophoto}

\end{document}